\documentclass[10pt,a4]{article}
\usepackage[top=0.85in,left=1.75in,footskip=0.65in,marginparwidth=0.5in]{geometry}

\usepackage[utf8]{inputenc}

\usepackage{natbib}

\usepackage{nameref,hyperref}

\usepackage[right]{lineno}

\usepackage{microtype}
\DisableLigatures[f]{encoding = *, family = * }

\raggedright
\usepackage{changepage}

\usepackage[aboveskip=1pt,labelfont=bf,labelsep=period,singlelinecheck=off]{caption}

\makeatletter
\renewcommand{\@biblabel}[1]{\quad#1.}
\makeatother

\usepackage{lastpage,fancyhdr,graphicx}
\usepackage{epstopdf}
\fancyheadoffset[L]{2.25in}
\fancyfootoffset[L]{2.25in}

\usepackage{color}

\definecolor{Gray}{gray}{.25}

\usepackage{graphicx}

\usepackage{sidecap}

\usepackage{wrapfig}
\usepackage[pscoord]{eso-pic}
\usepackage[fulladjust]{marginnote}
\reversemarginpar

\begin{document}
\vspace*{0.35in}

% title goes here:
\begin{flushleft}
{\Large
\textbf\newline{ \textbf{Tracking sustainability: co-evolution of economic and ecological activities in the industrialization of the United Kingdom and China}}
}
\newline
% authors go here:
\\
Xiaoyu Hou\textsuperscript{1},
Tianyi Zhou\textsuperscript{1},
Xianyuan Chang\textsuperscript{1},
Feng Mao\textsuperscript{2},
Zhaoping Wu\textsuperscript{1},
Ying Ge\textsuperscript{1},
Kang Hao Cheong\textsuperscript{3}
Jie Chang\textsuperscript{1,*}
Yong Min\textsuperscript{4,**}
\\
\bigskip
\b{1} College of Life Sciences, Zhejiang University, Hangzhou 310058, China
\\
\b{2} University of Warwick, Coventry, CV4 7AL, the United Kingdom 
\\
\b{3} School of Physical and Mathematical Sciences, Nanyang Technological University, 21 Nanyang Link, S637371, Singapore
\\
\b{4} Computational Communication Research Center, Beijing Normal University, Zhuhai, 519087, China 
\\
\bigskip
* Author for correspondence. E-mail: jchang@zju.edu.cn (CHANG J) or myong@bnu.edu.cn (MIN Y)

\section*{Abstract}
The co-evolution of economic and ecological activities represents one of the fundamental challenges in the realm of sustainable development. This study on the word trends in mainstream newspapers from the UK and China reveals that both early-industrialised countries and latecomers follow three modes of economic and ecological co-evolution. First, both economic and ecological words demonstrate an S-shaped growth trajectory, and the mode underscores the importance of information propagation, whilst also highlighting the crucial role of self-organisation in the accept society. Second, the co-occurrence of these two type words exhibits a Z-shaped relationship: for two-thirds of the observed period, they display synergistic interactions, while the remaining time shows trade-offs. Lastly, the words related to ecological degradation follow M-shaped trajectories in parallel with economic growth, suggesting periodic disruptions and reconstructions in their interrelationships. Our findings contribute to a more nuanced understanding of the co-evolutionary mechanisms that govern collective behaviours in human society.
 \end{flushleft}
 {Keywords:}
 Sustainability; culturomics; cultural co-evolution; social transition; newspaper content analysis; self-organization

\newpage
% the * after section prevents numbering
\section*{1 Introduction}
Long-term social trends and transitions driven by industrialization are among the primary concerns of sustainability science (\cite{Kates}; \cite{Chang}). The conflict and balance between economic activities and the environment have implications for human well-being (\cite{Palmer}; \cite{Frank}; \cite{Gu}). Various social transition models, including the environmental Kuznets curve and the Green Loop-Red Loop model, explore the potential for synergy or trade-offs, i.e., sustainability, between economic and ecological developments (\cite{Cumming2};\cite{Zhang2}; \cite{Anser}). However, a comprehensive empirical analysis on social transitions spanning the entirety of industrialization is lacking. The trajectory of social development presents alternating innovation trends of maintaining sustainability and facing collapse (\cite{Bettencourt}; \cite{Ali}). The intensity, trajectory, rates, transition points, interrelationships, and evolutionary endogenous mechanism of economic and ecological activities are pressing subjects given the rapid and significant transitions witnessed in many countries today.

Culturomics emerged a decade ago, offering a quantitative analysis of the development and evolution of human culture (\cite{Michel}; \cite{Lansdall-Welfare}; \cite{Hills}). It relies on extensive digital corpora, quantitative genetic principles and models, and evolutionary theory. Through this lens, the evolutionary trajectory of culture becomes evident. Topics of research include the relationships between individualism-collectivism and economic levels (\cite{Skrebyte}), the historical trends in national subjective well-being (\cite{Hills}), and the response of love in literary history to the economic level (\cite{Baumard}). A recent study noted that humans prefer innovative economic words and fixed ecological words (\cite{Zhang1}). The long-term tendency and evolution mechanism of social transition during the whole industrialization is a wider and deeper topic of culturomics.

The historical trends of economic and ecological activities during social transitions primarily rely on event analysis or chronicles (\cite{Liu1}; \cite{Boivin}). The trajectory of economic and ecological development within socio-ecosystems differ between the early industrialized countries, such as the UK, and latecomers, including Austria, New Zealand, Indonesia, and Myanmar (\cite{Krausmann}; \cite{Cumming1}). Some societies have transitioned towards ecological preferences or ecological friendliness along with economic growth (\cite{Liu};\cite{Leong}). A newspaper content analysis revealed that the societal value of water resources in Australia has aligned more closely with environmental-friendly awareness since 1981(\cite{Wei}). Studies have found that social transition presents wave mode (\cite{Schot}). However, these were not the results of culturomics studies. As of now, there is no quantitative conclusion regarding the rate and fluctuations in the relationship between economic and ecological activities or regarding the endogenous mechanism driving sustainability.

In this study, we chose the UK, an early industrialized country, and China, a latecomer, as case studies. We analysed the trends in economic and ecological activities and their relationship using mainstream newspapers as the corpus. Using the method and principles of culturomics, we studied the co-evolution tendency of economic and ecological activities in the two countries from the first industrial revolution (IR-I), second (IR-II), third (IR-III) to the fourth (IR-IV):the growth speeds of economic and ecological words; periods of fluctuation, their amplitude and variation; the direction of sustainability indicated by the relationship of ecological with economic words; the dynamics of economic and ecological words species, their diversity and turnover rate; the artificial selection or neutrality in evolution; and the trend of sustainability indicated by ecological degradation related words in response to economic levels of the societies.

\section*{2 Materials and Methods}

\subsection*{2.1 Data} 
\subsubsection*{2.1.1 Data of newspaper corpus} 
\subsubsection*{2.1.1.1 Create a target words library} 
To create a target word library that can comprehensively represent economic and ecological activities, we followed a ‘seven-step’ approach (\cite{Madin}) to construct the human economic activity ontology (HENAO) and the human ecological activity ontology (NELAO). Detailed ‘seven-step’ practice can be found in the Supplementary text 1.1.

We built the HENAO with three levels: the economic activities of a country into four major categories (primary classification) of activities: production, circulation, life-support, and regulation. Production activity includes agricultural activity, mining, manufacturing, and construction; circulation activity includes commercial activity, transportation, warehousing, and postal activity. The life-support activity includes accommodation and catering, education and healthcare, real estate, leasing, entertainment and leisure, and other activities. Regulatory activity includes financial and information activities. The above activities belonging to the second-level classification were further divided into the third level of classification; for example, agricultural activities can be further subdivided into planting activities, forestry activities, animal husbandry, and fishing. Manufacturing activities can be divided into 32 tertiary subcategories, including agricultural product manufacturing, food manufacturing, alcohol manufacturing, beverage manufacturing, and wood processing. We subdivided each second-level activity and ultimately obtained a total of 90 third-level classes (Supplementary Fig. 1).

From the three-level classification, we listed as many words as possible related to the processes and substances of each activity. For example, planting activities were subdivided into ‘planting’, ‘crops’, ‘fruits’, ‘harvesting’, ‘breeding’, ‘irrigation’, and ‘fertilization’.

We constructed the Human Ecological Activity Ontology (HELAO) with three levels by repeating the ‘seven-step’. We subdivided ecological activities into five subcategories at the first level: environmental improvement, pollution prevention, resource conservation, and sustainable activities and pollution activities. We re-subdivided a subcategory of positive and negative activities. For classification in the 2nd level, environmental improvement includes non-biological environmental improvement and biological-environmental improvement; the pollution prevention includes non-biological pollution prevention and biological pollution prevention; the resource conservation includes non-biological resource conservation, biological resource conservation, and energy-saving activities; the sustainable activity includes circular activities, renewable activities, green activities, etc.; the pollution activity includes non-biological environmental pollution and biological pollution, etc. Activities belonging to the second-level classification can be further subdivided, for example, non-biological environment improvement can be further subdivided into water environment improvements, atmospheric quality improvements, soil restoration, soil and water conservation, etc. Biological environmental improvement can be further subdivided into species protection, diversity protection, habitat protection, vegetation restoration, etc. In sequence, each secondary activity was subdivided to obtain a total of 57 tertiary classified activities (Supplementary Fig. 2). From the three-level classification of activities, we listed as many words as possible related to the processes and substances of each activity, such as water environment improvement, which can be subdivided into water environment, water quality, and water restoration.

Additionally, we supplemented the concepts in Step (2) of the seven-steps method (see Supplementary text 1.1) with statistical data from the British National corpus (BNC, http://www.natcorp.ox.ac.uk/) and Beijing Language and Culture University corpus (BCC, http://bcc.blcu.edu.cn/) for specific “n-grams”, which are sequences of terms separated by a space. For example, ‘ecosystem’ is a 1-gram, and ‘ecological protection’ is a 2-gram (hereafter, we denote n-gram terms in italics). The concepts are found following the method of ‘fill-in-the-blank’ (\cite {Anderson} et al., 2021). For example, when we searched for the most common 1-gram after environmental, we thought environmental protection would be one possible answer, and we referred to protection as a “blank” term. We focused on the 1-gram and 2-gram datasets and ensured that the words filling in the blank space were at least four characters long. The selection of economic and ecological core words was based on the word wants reverse dictionary system (produced by the natural language processing laboratory of Tsinghua University) to query the possible econometric and ecological synonyms (Supplementary Table 1). For example, the core words of economic words are economists (economy, economic), finance (finance, financial), comers (commerce, commercial), etc.

Iterating the above steps, we ultimately extracted nearly 3,000 target words related to human economic activities and nearly 1,000 target words related to ecological activities.

\subsubsection*{2.1.1.2 Data acquisition and filtering} 
We collected the occurrence of words in the newspapers over the years according to the targeted word library. For the data source representing the UK, we choose The Times, which has always been regarded as the first mainstream newspaper in the UK and is known as "the faithful recorder of society". The occurrence of words from 1785 to 2014 comes from Gale: The Times digital archive corpus (https://go.gale.com/). For China, an early data source is Shun Pao, a widespread newspaper in China before the establishment of People’s Republic of China in 1949. After 1946, the data source was changed to The People's Daily, a mainstream newspaper published by the central government. The words from 1872 to 2015 come from the Beijing Language and Culture University corpus (BCC, http://bcc.blcu.edu.cn/, \cite{Xun})

We focused on 1- and 2-grams because the most valuable common words that concern us comprise two words or less, and the computational burden increases significantly with the increase in n-gram length. Based on HENAO and HELAO, before the final analysis, our preliminary analysis revealed some necessary data-cleaning steps. (1) The n-grams with a total occurrence of less than 8 in the full-text corpus were removed. (2) All n-grams containing numbers and punctuation, all 1-grams containing only three or fewer characters, and all 2-grams containing three or fewer characters in the first or second term, such as green @ and c economy, were deleted. (3) For the 2-gram, only the phrases containing adjectives and nouns were retained, such as for ecology, were deleted. (4) The singular and plural words (such as environment and environments), and abbreviations (such as GDP and gross domestic product and its plural form) were combined. We retained 2,225 economic n-grams and 720 ecological n-grams in The Times of the UK and 3,347 economic n-grams and 1,249 ecological n-grams in Shun Pao and The People’s Daily of China.

We use the Gale The Times Digital Archive 1785-1985 Corpus to obtain the occurrence of target n-grams in The Times. When the target n-gram is counted once, repeat occurrences of the n-gram in the same article will not be counted again, which may cause the data to be underestimated. To correct this deviation, we use a sampling survey to calculate statistics of the probability of different n-grams appearing in the same article, obtain correction coefficients to correct the primary data, and then calculate an annual occurrence of n-grams. To obtain the occurrences of an n-gram in each article, we first randomly extract the target n-gram in groups. After the economic and ecological n-grams are arranged in descending order of word frequency, they are divided into 5 groups. Ten n-grams from each group are randomized with a total of 50 words to cover n-grams at all levels. Then, we use visual interpretation to obtain the repeat occurrences of an n-gram in an article in a certain year. We still use the random program to select the year to be interpreted, with a total of 10 years selected. We randomly selected 10 articles per year and manually checked 3,000 articles in total.

We used random sampling to calculate the average number of words (\textit{ \({N}_{i}\)}) in one article (\cite{Wei}). The method of sampling is as follows: first, we randomly sampled the years, i.e., during 1780-2014, we randomly selected 3 years to read at an interval of 40 years, for a total of 18 sample years. Second, we randomly selected 3 months in each selected year and randomly selected 3 dates in each selected month. Then, we read 162 newspapers and more than 21,000 news reports and tabulated the word count of each article. We use the Gale The Times Digital Archive 1785-1985 Corpus to obtain the average frequency of “the”, “and” and “a” as the number of articles in the current year (\textit{\({A}_{i}\)}) . Then, the annual word occurrence (\textit{\(\bar{W}\)}) of The Times,

\begin{equation}
    {N}_{i}= {A}_{i}\times \bar{W}
\end{equation}

The occurrence of n-gram in China we obtained in BCC corpus was not underestimated, the annual word occurrence (\textit{\({N}_{i}\) }) of Shun Pao and The people’s daily comes from BCC corpus (\cite{Zhang1}; \cite{Xun}).

\subsubsection*{2.1.2 Social economic data} 

GDP per capita from 1960 to 2020 for the UK and China comes from the World Bank (https://data.worldbank.org/).

\subsection*{2.2 Division of the periods of the industrial revolution} 

The division of the four industrial revolutions can be found in Supplementary Text 1.2 and supplementary table 2.

\subsection*{2.3 Calculations} 
\subsubsection*{2.3.1 Occurrence and frequency of words} 

The occurrence of words in newspapers is affected by “survivor bias” (\cite{Miller}). That is, due to page limitations, the information appearing in newspapers has been filtered according to preference intensity, and it may deviate from the actual situation. The frequency \textit{\({P}_{t}\)}, standardized by formula (2), following the method used by \cite{Michel}, serves as a proxy for economic and ecological activities competing with other content in limited space and eliminates the impact of newspaper space (total word number) on the occurrence of words, as in the following equation:

\begin{equation}
    {P}_{t}= \frac{{O}_{t}}{{N}_{t}}
\end{equation}
 where \textit{\({P}_{t}\)} is the occurrence of a target n-gram at year \textit{t} and\textit{ \({N}_{t}\)} is the occurrence of all words in the newspaper within a year.
 
\subsubsection*{2.3.2 Period and amplitude of fluctuations} 

We marked the corresponding years for the peak and valley values of the occurrence of words (Supplementary Table 3), with the year between the two valley values ( \textit{\({O}_{valley1}\) }and \textit{\({O}_{valley2}\)}) as one fluctuation period (\textit{\({T}_{c}\)}). The amplitude before the peak (\textit{\({O}_{peak}\)}) is shown in formula (3) and another amplitude after the peak is shown in formula (4),
\begin{equation}
   {A}_{i1}={O}_{peak}-{O}_{valley1}
\end{equation}

\begin{equation}
   {A}_{i2}={O}_{peak}-{O}_{valley2}
\end{equation}
where \textit{\({A}_{i1}\) }and \textit{\({A}_{i2}\)} are the two amplitudes of a fluctuation within a certain period, respectively; . The criteria for defining peak and valley values are that the absolute value of the ratio of a certain amplitude to the adjacent previous valley or peak value is higher than 5\% . 

\subsubsection*{2.3.3 Coefficient of variation}

The standard deviation (\textit{SD}) of word occurrence changes within a fluctuation is
\begin{equation}
   SD=\sqrt{\frac{\sum_{i=1}^{n}{{O}_{i}-\bar{O}}}{n-1}}
\end{equation}
where \textit{\({O}_{i}\) }is the occurrence of the target words of the fluctuation, \textit{\(\bar{O}\) }is the average occurrence of words within the fluctuation, n is the number of years within a fluctuation.

We used the coefficient of variation (\textit{CV}) to measure the fluctuation of the occurrence of words,
\begin{equation}
CV(\% )= \frac{SD}{\bar{O}}\times100
\end{equation}
where \textbf{SD} is calculated using formula (5),\textit{ \(\bar{O}\) }is the average occurrence of words within the fluctuation.

\subsubsection*{2.3.4 Relative abundance}

The frequency of words decreases as they become passé in public discourse due to their widespread use (\cite{Bentley}). In other words, the words are cooling--- a reduction of marginal effects of the familiar activities. Therefore, we used the relative abundance to measure the preferences for different kinds of words, as:
\begin{equation}
{R}_{(t)}= \frac{{O}_{i(t)}}{\sum_{i=1}^{n}{O}_{i(t)}}
\end{equation}
where\textit{\({O}_{(t)}\)}is the occurrence of a target word in year \textit{t}.

\subsubsection*{2.3.5 Diversity of word species}

The Simpson index is a comprehensive index of richness and evenness and is usually used to evaluate species diversity (\cite{Ouyang}). The Simpson index is calculated as:
\begin{equation}
S= 1- \sum_{i=1}^{s}{k}_{i}^2
\end{equation}
where \textit{S} is the number of word species and \textit{\({K}_{i}\)} is the proportion of the number of individuals of species \textit{i} relative to the total number of individuals in the community.

\subsubsection*{2.3.6 The distribution of words}
The power law distribution between the proportion and rank can reflect the degree of dominance of a particular species in a population, which can be expressed as:
\begin{equation}
{K}_{i}= c{R}^{-\alpha }
\end{equation}

where \({K}_{i}\) is the same as formula (8), R is the ranking of word species in descending order of its number of individuals, and \textit{\(c\)} is a constant. When \textit{\(\alpha\)}  approaches 1, the word species follows the Zipf distribution, which means that the evolution of word species is driven by random drift.

\subsubsection*{2.3.7 Turnover rate and turnover distribution}

The turnover rate is used to calculate the variation in species composition among sites (\cite{Baselga}). We consider all the word species of a type of word (such as economic words) within a year as a site. The turnover rate (\textit{Z}, word species \(yr^{-1}\) )can be used to represent the variation speed in word species combinations between two sites; it is calculated as:
\begin{equation}
Z= \frac{{S}_{u}}{{T}_{int}}
\end{equation}

where \textit{\(S_{u}\)} is the sum of the number of species unique to the two sites and \textit{\(T_{int}\)} is the time interval between the two sites for comparison.

We define the turnover distribution by using the method by \cite{Acerbi} based on the changes generated by the Wright-Fisher model, which is a Markov process. In the list of the word species ranked in descending order of its number of individuals with y as the capacity, a list is a site, the \textit{Z} of the list is the same as formula (10), and \textit{\(T_{int}\)} is one year. The top list sizes y in this study were from the \(1^{st}\) to the \(30^{th}\). A turnover distribution is an Power regression on 30 pairs of \textit{Z }and \textit{y} in a \textit{\(T_{int}\)}, as shown in the following formula:

\begin{equation}
Z = e{y}^{b}
\end{equation}

where \textit{b} is the turnover exponent and e is a constant.

The index b determines the shape of the turnover distribution. According to \cite{Acerbi}, in the case of the neutral model, \textit{b} is 0.86, which means that the traits are randomly copied. When \(\textit{b} > 0.86\), the selection fixes the high-frequency traits and is called ‘positive frequency-dependent’. In contrast, if \(\textit{b} < 0.86\), there is a negative selection for high-frequency traits, and then the popular trending becomes decentralized over time, called a negative frequency-dependent (Fig. 1).

\begin{figure}[ht] %s state preferences regarding figure placement here
\centering
% use to correct figure counter if necessary
%\renewcommand{\thefigure}{1}

\includegraphics[width=\textwidth]{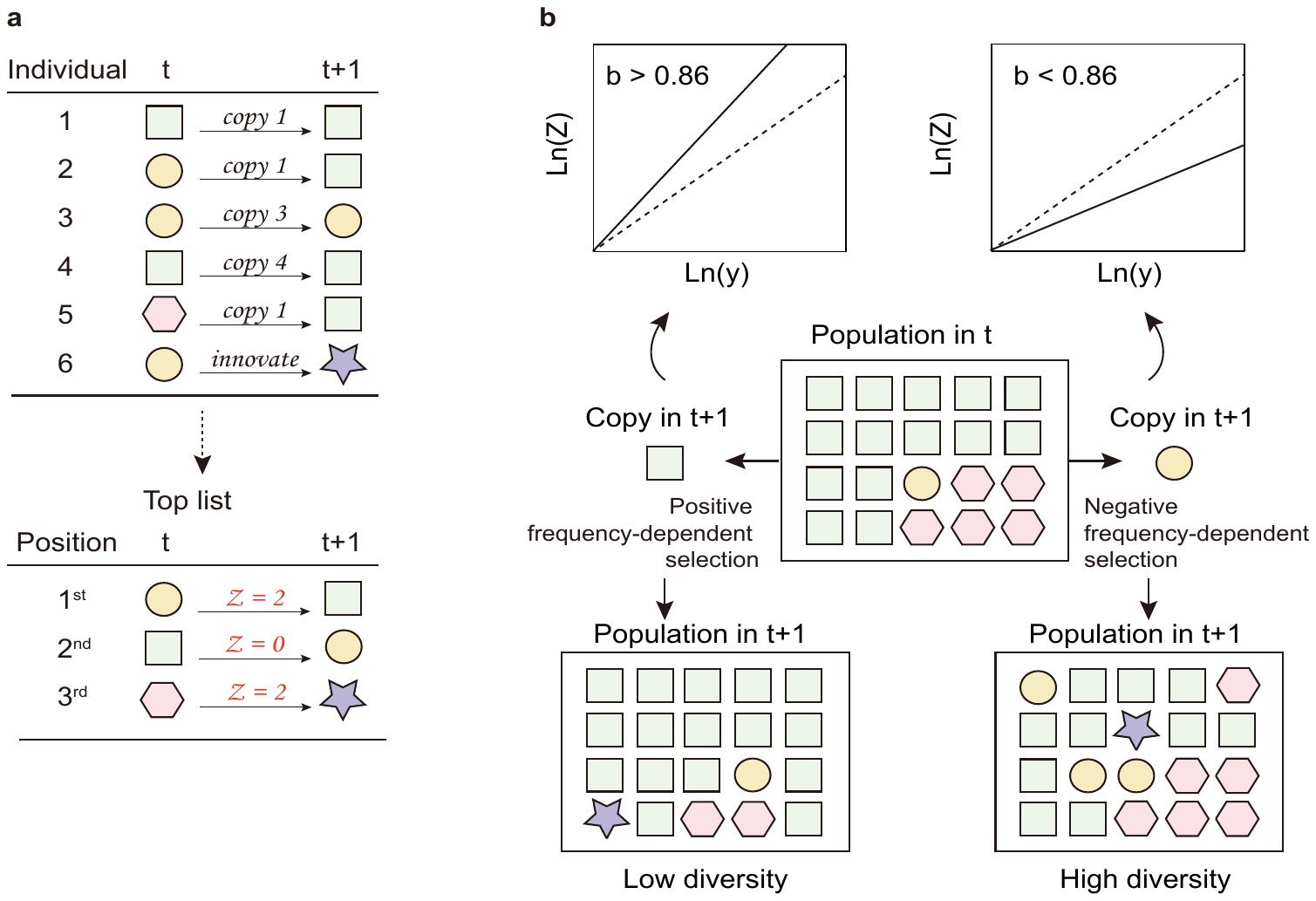}

\caption{\color{black} \textbf{Schematic diagram of turnover distribution and selection.}. \textbf{ a,} {Distribution of six individuals and the turnover rate of the top 1 to top 3 lists at two time points (when \textit{y} = 1, \textit{Z} = 2; when \textit{y} = 2, \textit{Z} = 0; and when \textit{y} = 3, \textit{Z} = 2). \textbf{b,} Different types of selection will form different turnover curves (\textit{\( z = ey^b\)}). When a larger number of individuals (high-frequency traits) are selected, i.e., positive frequency dependent selection, the turnover index \textit{b} is higher than 0.86 (dotted line), and the dominant individuals are fixed, resulting in a decrease in diversity (evenness). When high-frequency traits are negatively selected, \textit{b} is below 0.86 (dotted line), and the dominant individuals are replaced rapidly, leading to an increase in diversity.}}

\label{fig1} % \label works only AFTER \caption within figure environment

\end{figure}

All parameters in this paper in this paper see Supplementary Table 4.

\subsubsection*{2.3.8 Statistical analysis}

We use linear fitting (including the univariate quadratic equation and univariate quadratic equation) to obtain the correlation coefficient between the turnover rate and interval years between the occurrence of words and economic level. We use power-law fitting to obtain the correlation coefficient between the fluctuation period and time, between the proportion of word species and rank, and between the turnover rate and top list size. The analyses are performed in SPSS 20 (SPSS Inc., Chicago, IL, USA).

Nonlinear fitting analysis is conducted for two sets of word frequencies (\cite{Wei}). The fitted sigmoid curves for the word frequency  \textit{\({P}_{t}\)} of economic activities and ecological activities against time  \textit{\(t\)} are described by
\begin{equation}
{P}_{t}=\frac{{a}_{1}-{a}_{2}}{1+{(\frac{t}{{t}_{0}})}^{\beta }}+{a}_{2}
\end{equation}

The regression values of \textit{\(a_1\)}, \textit{\(a_2\)}, \textit{\(t_0\)}, and \textit{\(\beta\)} are listed in Supplementary Table 5. We use the nonlinear fitting module in Origin (Origin 2021, Originab Corporation) to check whether the model coefficients are statistically significant.
Then, we use the first derivative of the fitted curve to calculate the growth rate of word frequency and use the inflection point of the second derivative to divide the different stages of the S-mode.

% newpage forces a page break if you want to clearly separate materials from results
\newpage

\section*{3 Results}
\subsection*{3.1 Economic and ecological activities grow in fluctuations}

The Times, a mainstream UK newspaper established in 1785, has a total word count of \(8.8\times 10^9\) over the past 230 years. In the UK, the frequency of economic words in IR-I (1785-1865), IR-II (1866-1945), IR-III (1946-1992), and IR-IV (1993-2014) has gone through an S-shaped curve, which can be divided into the stages of accumulation, acceleration, stabilization, and decline (Fig. 2a). The frequency reflects the intensity of change in economic or ecological traits that people are concerned about timely, given that news reflects new social activities or changes in existing things, and the limited spaces in newspapers. The decline in the frequency of economic words after 1978 indicated that economic activities had entered the stabilization stage of the logistic curve (Supplementary Fig. 4). The frequency of ecological words began to accelerate in the 1950s, far behind economic words and was still growing.

\begin{figure}[ht] %s state preferences regarding figure placement here

% use to correct figure counter if necessary
%\renewcommand{\thefigure}{2}

\includegraphics[width=\textwidth]{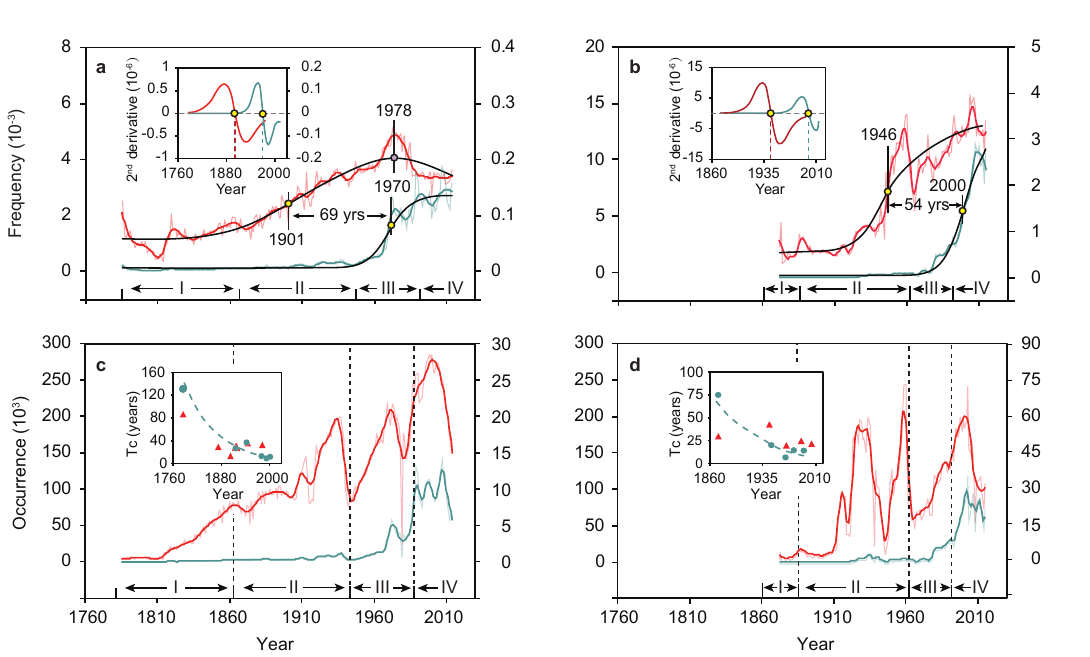}

\caption{\color{Gray} \textbf{Historical trends of the frequency and occurrence of economic and ecological words.} Red represents economic words, and green represents ecological words. \textbf{a-b}, Frequency of words: the UK (\textbf{a}), China (\textbf{b}), inserts show the second derivative of S-mode, the annotation year represents the turning point of maximum acceleration (yellow dots) and decline (purple dot) at stages of S-mode.\textbf{c-d},Occurrence of words: the UK (\textbf{c}), China (\textbf{d}), insert shows the fluctuation cycles (\textit{\({T}_{c}\)}) decreasing over time, with triangles representing economic and circles representing ecological words. The trends were smoothed by the Loess method with a window rate of 10\%,  and thin lines represent the actual word data. In the X-axis, I, II, III, and IV represent the four industrial revolutions (Supplementary Table 2). }

\label{fig2} % \label works only AFTER \caption within figure environment

\end{figure}

China does not have a complete industrial history, as it started 100 years later than the UK. The growth of economic frequency experienced a shorter and steeper acceleration stage. The growth rate was the largest in 1946 before New China was established (Supplementary Fig. 4). The frequency increased to a peak when the Great Leap Forward (1958-1960) occurred and then sharply declined during the Cultural Revolution (1966-1976). Afterwards, it again increased and was still increasing (Fig. 2b). This indicates that economic activities have not yet entered the stabilization stage. The maximum acceleration stage of ecological activities in China started 30 years later than that in the UK, and the lag time between ecological and economic maximum acceleration in China was nearly 20 years shorter than that in the UK. Moreover, the frequencies of words in China were higher than those in the UK. Notably, no significant punctuated changes in this process of two kinds of activities were observed by any IRs, either in the UK or China.

Word frequency eliminates the impact of total word size on newspapers (Supplementary Fig. 5), but it hides some details of the fluctuations in word occurrence itself. In the UK, the occurrence of economic words increased from 1810 and decreased after reaching its peak in 2000 with six fluctuations. The first fluctuation lasted for 87 years and then continuously shortened and stabilized for 32 years (Fig. 2c and insert; Supplementary Table 3). The occurrence of ecological words was relatively low and increased slowly before 1960, and then rapidly increased until 2010 with six fluctuations. The fluctuation periods also shortened overtime recently and stabilized to 11 years. Notably, it conformed to the power equation (\({T}_{c}=a{y}^{-\beta })\) with super-linearly attenuation \(\beta\) = 22. The ratio of ecological word occurrences to economic ones has gradually increased from 0.5\% to 5.2\% in the past 230 years. The growth of the occurrence of economic words in China occurred later but was sharp and had massive fluctuations. The periods also shortened; the last period was 22 years (Fig. 2d and insert; Supplementary Table 3). The occurrence of ecological words in China was typically low before 1972 and then increased sharply. The periods of fluctuations also shortened over time and conformed to the power equation with \(\beta\)= 30; the recent period was 11 years. Notably, the fluctuations in the two countries were unrelated to the IR transit points but related to events such as world wars, the oil crisis, and environmental conferences.

In the UK, the fluctuation amplitude of the occurrence of economic words expanded after World War I. The amplitude of ecological words was approximately one-tenth of that of economic words and expanded after World War II (Fig. 3a). In China, significant fluctuations in the occurrence of economic words began in the 1900s (Fig. 3b). The amplitude of ecological words expanded from the 1970s and was equivalent to one-third of economic words. In contrast to the amplitude, the coefficient of variation (CV) for economic words was lower than that for ecological words, and the CVs for these fluctuations decreased over time and stabilized similarly at about 20 in the UK (Fig. 3c; Supplementary Table 3). In China, the CV of the fluctuations in economic words was about 20\%, while that of ecological words was 10\% recently (Fig. 3d; Supplementary Table 3).

\begin{figure}[ht] %s state preferences regarding figure placement here

% use to correct figure counter if necessary
%\renewcommand{\thefigure}{3}

\includegraphics[width=\textwidth]{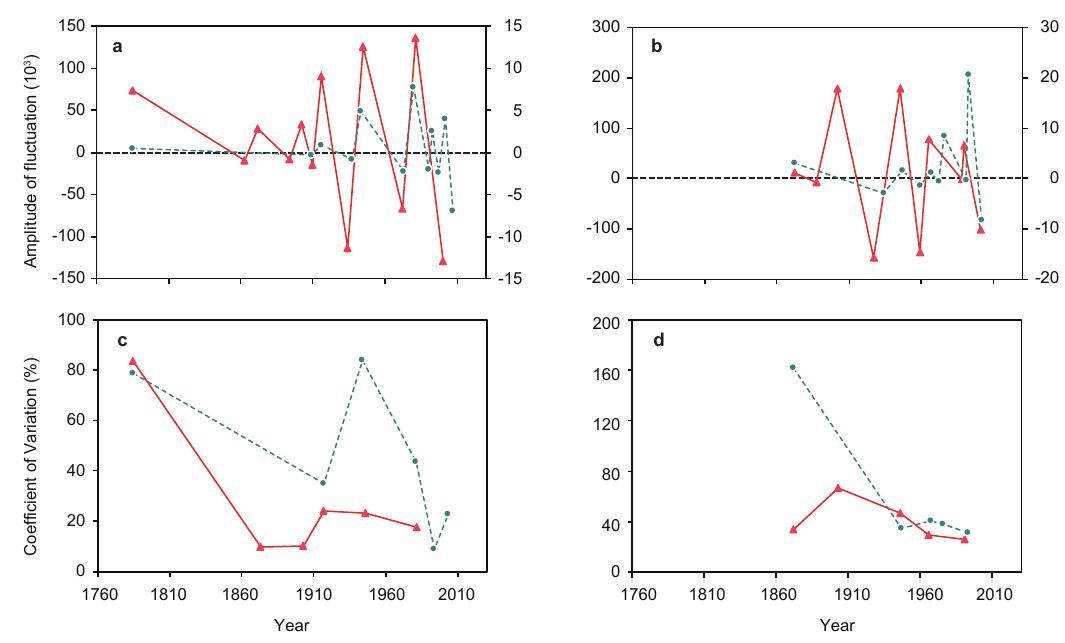}

\caption{\color{Gray} \textbf{Variations in the occurrence fluctuations of economic and ecological words in the UK (a, c) and China (b, d)}  Red triangles represent economic words; green circles represent ecological words. The points correspond to the year when each fluctuation starts. \textbf{a-b,} Amplitude of economic words (using primary coordinates) and ecological words (using secondary coordinates).\textbf{ c-d,} Coefficient of variation (CV) of economic and ecological words.}

\label{fig3} % \label works only AFTER \caption within figure environment

\end{figure}

\subsection*{3.2 Synergistic growth and trade-offs between economic and ecological activities}

The trends in the occurrence of economic and ecological words synergistically increased before 2007 in the UK, and then a dual recession of both economic and ecological words occurred. Previously, there were also dual recessions in 1938 during World War II and 1974 during the economic recession. It presented a clear zigzag (Z-shaped trajectory) with dual increasing, individual increasing, and dual declining. From 1981 to 2007, there was a dual increase overall with several fluctuations in 1990, 1993, 1997, and 2001 during when there were individual increases or decreases of economic or ecological words (Fig. 4a). In China, the fluctuations were simpler than those in the UK, though a clear Z-shaped trajectory can also be seen. Before 1970, the occurrence of ecological words remained very low and declined beginning in 1958. From 1964 to 2001, a long-term dual increase and then a dual decline to 2007 occurred. From 2008 to 2011, ecological words increased, while economic words decreased. Since then, ecological terms have decreased rapidly (Fig. 4b). Notably, although the national conditions of the UK and China are different, they both experienced turning points in 2001 (Supplementary Table 6).

\begin{figure}[ht] %s state preferences regarding figure placement here

% use to correct figure counter if necessary
%\renewcommand{\thefigure}{4}

\includegraphics[width=\textwidth]{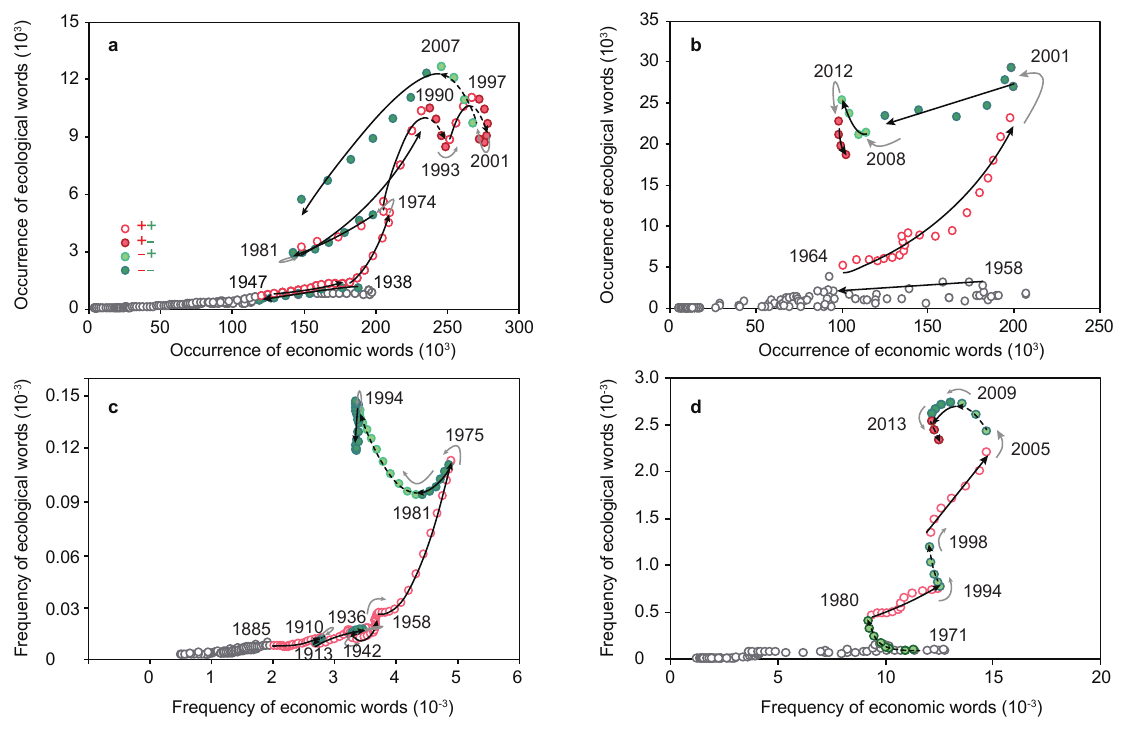}

\caption{\color{Gray} \textbf{Relationships between economic activities and ecological activities in the UK (a, c) and China (b, d). a-b,} Occurrence of words;\textbf{ c-d,} frequency of words. The arrow indicates the direction in the time series. The annotation year is the starting year of changes. For the trend in the early period (grey dots), see Supplementary Fig. 6.}

\label{fig4} % \label works only AFTER \caption within figure environment

\end{figure}

The frequencies of words reveal a clearer trajectory than occurrences. In the UK, overall, the trend of ecological frequency was also a Z-shaped curve relative to the economic word frequency (Fig. 4c). Ecological words began to increase rapidly in the early 20th century and increased exponentially relative to economic words until 1975, at which point there was no further dual increase. From 1975 to 1981, the frequency of economic words decreased, while that of ecological words increased until 1993. After 1994, both words plummeted. A similar Z-shaped trend was observed in China (Fig. 4d). The start of a dual increase was delayed until the 1980s, before which there was a 10-year reversal. In 1994 and 2005, there was also a trade-off where economic words decreased and ecological words increased. After 2009, there were two more reversals.

\subsection*{3.3 Dynamics of word abundance, diversity, and turnover rate}

In the UK, among 11 major categories of economic words, the relative abundance of catering-related words belonging to the ‘life support type’ was highest at the beginning of industrialization. The financial-related words replaced it after 1800 and remained predominant until the present (Fig. 5a). During IR-III, the relative abundance of manufacturing-related words increased but decreased in IR-IV when the abundance of information dissemination-related words increased. Among the 5 major categories of ecological words, pollution-related words had a high relative abundance, followed by energy-saving-related words in IR-I and II (Fig. 5b). From the 1900s (during IR-II), the relative abundance of environmental protection-related words began to increase and has held the top ranking since the 1920s.

\begin{figure}[ht] %s state preferences regarding figure placement here

% use to correct figure counter if necessary
%\renewcommand{\thefigure}{5}

\includegraphics[width=\textwidth]{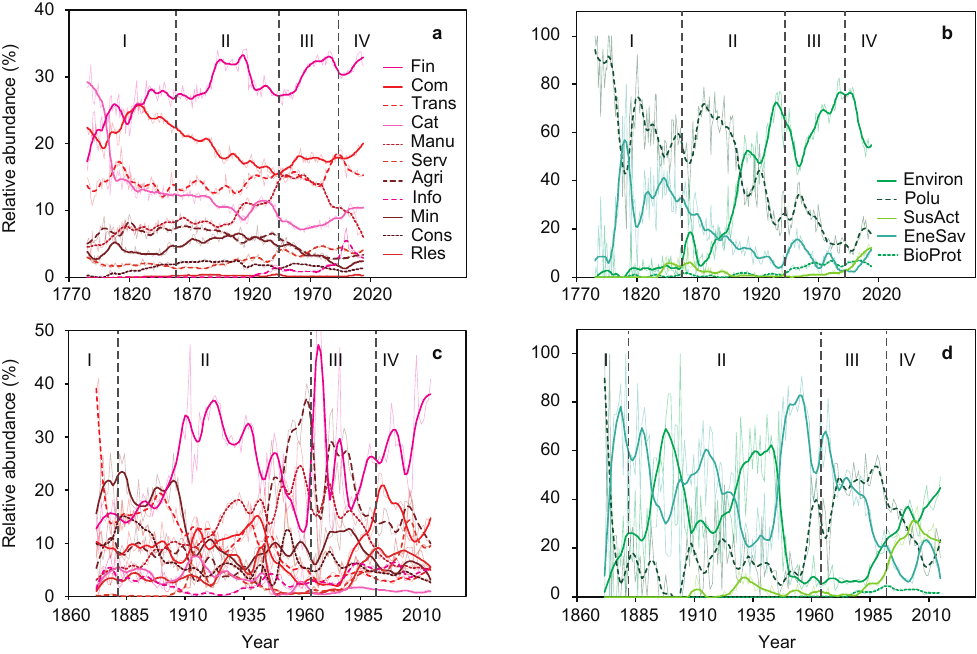}

\caption{\color{Gray} \textbf{Relative abundance of the categories of economic and ecological words in the UK (a-b) and China (c-d).} It was smoothed by the Loess method with a window rate of 10\%; thin lines represent the actual data. Red represents economic, and green represents ecological words. Abbreviations: Fin-financial, Com-commercial, Trans-transportation, Cat-catering, Manu-manufacture, Serv-serving, Agri-agricultural, Info-information, Min-mining, Cons-construction, Rles-real estate; Environ-environmental promotion, Polu-pollution, SusAct-sustainable activity, EneSav-energy-saving, BioProt-biological protection.}

\label{fig5} % \label works only AFTER \caption within figure environment

\end{figure}

In China, the relative abundance of words related to transportation, mining, and construction ranked high until 1900, dissimilar to that in the UK (Fig. 5c). After 1910, financial-related words increased to the top and remained until recent years. The ecological words also differed from those in the UK. Environmental protection- and energy-saving-related words held high relative abundance until 1960, followed by pollution-related words, which have increased since IR-III (Fig. 5d). Recently, the relative abundance of words related to environmental protection and sustainable activities has increased.

The diversity trends of the word species indicated by Simpson index (S), which integrates species richness and evenness performed that, the S of economic had a much higher value (0.95) than that of ecological (0.7) in the early stage of industrialization in the UK (Supplementary Table 7). However, the two have converged to about 0.96 in recent years (Fig. 6a). Similarly, in China, the S of economic (0.91) was also much higher than the ecological S (0.28) in the early stage, and they converged to about 0.95 recently (Fig. 6b; Supplementary Table 7). The S of ecological word species in the UK reached about 0.9 as early as 1850, while in China, it did not reach this level until 1950, with a time lag of almost 100 years.

\begin{figure}[ht] %s state preferences regarding figure placement here

% use to correct figure counter if necessary
%\renewcommand{\thefigure}{6}

\includegraphics[width=\textwidth]{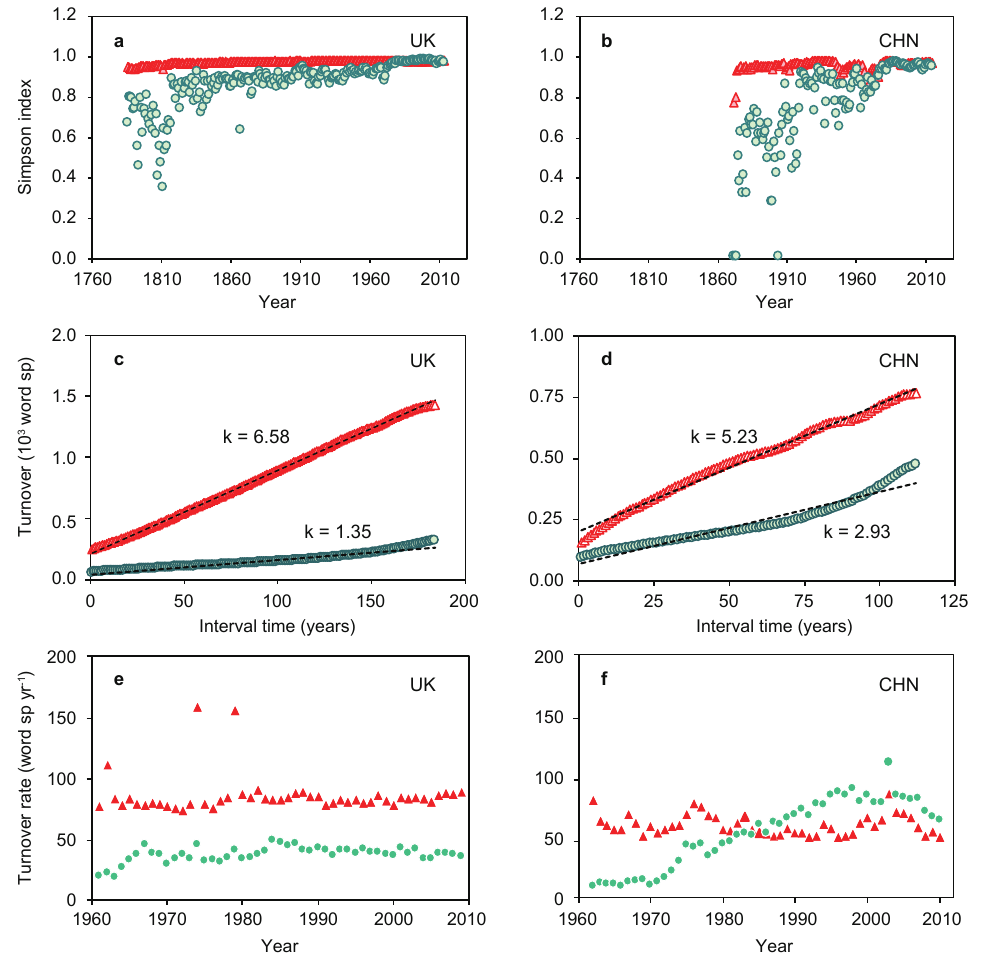}

\caption{\color{Gray} \textbf{Historical trend of word species change.}  \textbf{a-f,} Trends of diversity index (\textbf{a-b}), turnover rate along with the interval years’ increase (\textbf{c-d}), turnover rate of economic and ecological words from 1960 to 2010 (\textbf{e-f}), Red triangles represent economic words; green circles represent ecological words.}

\label{fig6} % \label works only AFTER \caption within figure environment

\end{figure}

The turnover of word species (sp) increased linearly with the increase in the interval between comparative years in both countries. In the UK, the turnover number of economic words increased at a constant rate (\textit{k}) as the interval time of newspaper publication increased (Fig. 6c), similar to a molecular clock. The turnover of ecological word species almost linearly increased with the extension of the interval year, but the \textit{k} was only 1/5 of that of economic word species. Similarly, in China, there also a constant rate of economic word species (Fig. 6d), while the \textit{k} is 4/5 that in the UK. In contrast, the \textit{k} of ecological word species is 2.2-fold that in the UK. Since 1960, in the UK, the turnover rate has stabilized at about 80 sp  \(yr^{-1}\) for economic and about 50 sp  \(yr^{-1}\) for ecological (Fig. 6e). In China, the turnover rate for economic word species was lower than that in the UK and stabilized at about 60 sp \(yr^{-1}\), while that for ecological word species has not yet reached a stable rate (Fig. 6f). 

\subsection*{3.4 Evolutionary mechanism of economic and ecological activities}

Both economic and ecological activities followed a power law (\(K=c{R}^{-\alpha}\)) in the UK (Fig. 7a, b). After 1960, the high-frequency economic word species that were strongly artificially selected for \(\alpha\) was much higher than 1 (Fig. 7a). The judgment criterion is if the word species followed the Zipf distribution, i.e., \(\alpha\) close to 1, the turnover of words is a random copy. The high-frequency ecological words were also artificially selected for \(\alpha\) was also higher than 1 (Fig. 7b). In China, the \(\alpha\) values for the two kinds of word species were both higher than 1, and the \(\alpha\) of economic word species was also higher than that of ecological ones (Fig. 7c, d). The \(\alpha \)values for the two kinds of word species in China were slightly lower than those in the UK. This indicates that the selection for economic activities was stronger than that for ecological activities, and the selection in the UK was stronger than that in China.

\begin{figure}[ht] %s state preferences regarding figure placement here

% use to correct figure counter if necessary
%\renewcommand{\thefigure}{7}

\includegraphics[width=\textwidth]{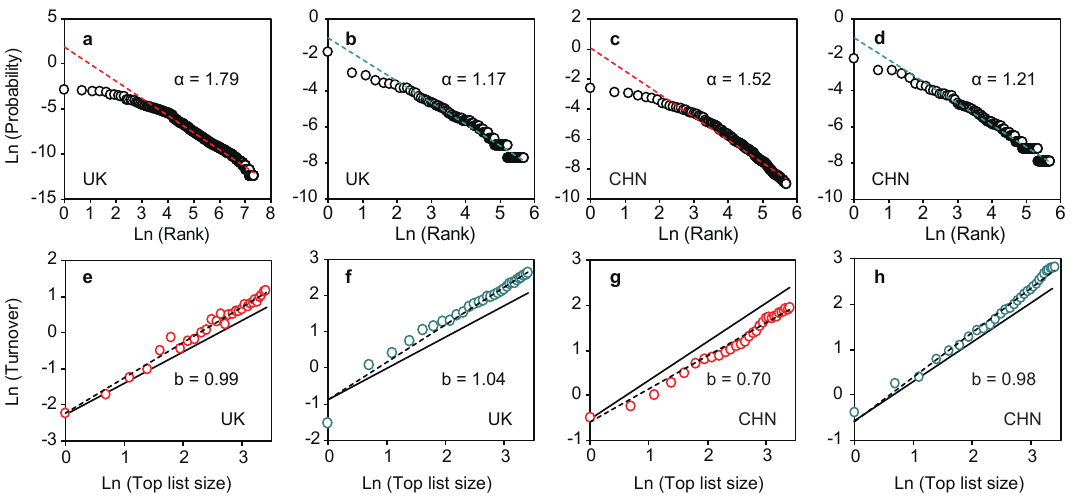}

\caption{\color{Gray} \textbf{Distributions of economic and ecological word species in the UK (a-b, e-f) and China (c-d, g-h).} Red represents economic words, green represents ecological words. a-d, Power law. \textbf{e-h}, Turnover distribution, solid line represents a neutral value of 0.86.}

\label{fig7} % \label works only AFTER \caption within figure environment

\end{figure}

Due to the deviation of the distribution at the top from the power law, these values were determined by another selection model. In this model, if the turnover index b was near 0.86 (\(Z=e{Y}^{b}\)), a random drift without selection occurred. After 1960, in the UK, for the top 30 list economic and ecological word species, the b is higher than 0.86 (Fig. 7e, f), indicating that the high-frequency word species were fixed by selection, as evidenced by the slower turnover in the higher position of the list. In contrast, in China, after 1960, the b is lower than 0.86 for economic word species (Fig. 7g), indicating that selection accelerates the change in the higher position of the word list and that the high-frequency word species change quickly under selection. For ecological word species, the b that higher than 0.86 indicates that high-frequency words were fixed by selection (Fig. 7h).

%\clearpage makes sure that all above content is printed at this point and does not invade into the upcoming content
%\clearpage

\subsection*{3.5 Multiplicity of ecological degradation and improvements along with economic development}

The negative ecological words, which are related to environmental pollution and ecological decline, and the occurrence of negative ecological words had 4 fluctuations in response to GDP per capita from 1960 to 2014 in the UK (Fig. 8a). Each fluctuation showed a quadratic function (Supplementary Table 8). In China, the occurrence of negative ecological words exhibited 2 inverted U-shaped quadratic functions in response to the GDP per capita from 1960 to 2015 (Fig. 8b). For the ratio of ecological-to-economic words, there were 4 quadratic functions in response to GDP per capita from 1960 to 2014 in the UK (Fig. 8c). In China, there was only one inverted U-curve in response to GDP per capita (\(p < 0.05\), Fig. 8d), similar to the first inverted U-curve in the UK.

\begin{figure}[ht] %s state preferences regarding figure placement here

% use to correct figure counter if necessary
%\renewcommand{\thefigure}{8}

\includegraphics[width=\textwidth]{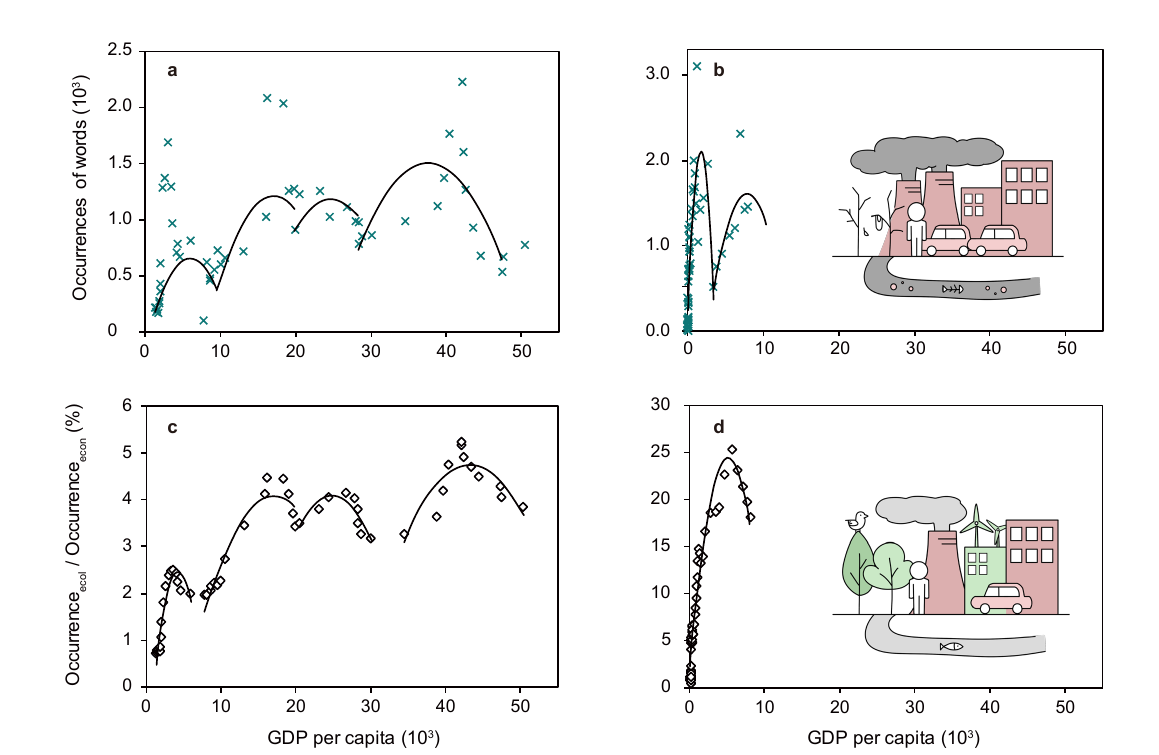}

\caption{\color{Gray} \textbf{Response of negative ecological word occurrence (a-b) and ecological-to-economic word ratio (c-d) to GDP per capita from 1960 to 2020.} The UK (\textbf{a, c}), China (\textbf{b, d}). Cross represents negative ecological words, and diamond represents the ratio of ecological words to economic words.}

\label{fig8} % \label works only AFTER \caption within figure environment

\end{figure}

\section*{4 Discussion}
\subsection*{4.1 An endogenous growth model}

The S mode, which can be described by the logistic model, is a general self-organizing process in the growth of organisms, population size, species accounting for evolution, and innovations and value changes of human society (\cite{Wei};\cite{Markard}; \cite{Mao}; \cite{Ali}). The increasing and decreasing economic words in the UK indicate that economic activities have accelerated since the 1870s and reached a stable stage of the S mode since 1978. Similarly, the economic words show that economic activities in China have also experienced a stage of accumulation and acceleration. The findings support the hypothesis that the development of industrialized societies is similar to that of biological embryos (\cite{Wen}). Of course, the development path of the latecomer industrializing countries may be different from that of early industrial countries (\cite{Krausmann}; \cite{Cumming1}). It suggests that there is a general S mode for social transition by industrialization, although the details of implementation may vary.

Ecological activities and consciousness are regulatory factors in the co-evolution between human society and nature for sustainability (\cite{Chang}; \cite{Palmer}). The dynamics of the words in the UK and China support the understanding that public ecological awareness lags behind economic development (\cite{Liu1}). Since the 1960s, the growth of ecological words in the UK has accelerated, but the proportion of ecological words in economic words has been only 5\%  until recent years. This indicates that the sustainability of industrialized countries is still low. The time lag between ecological and economic activities in China is shorter than that in the UK, although the acceleration of ecological words occurred 20 years later. This indicates that the cultural dissemination effect accelerates social transition (\cite{Liu1}; \cite{Meng}). In China, the proportion of ecological to economic words has reached about 25\% , which is higher than that in the UK. Notably, the existence of the accumulation stage of both economic and ecological words in China also indicates that dissemination can only promote but not replace societal self-development of the society being disseminated.

The four Industrial Revolutions have been found to influence the replacement of social terms related to health and the environment (\cite{O'Brien}; \cite{Atenstaedt}). Our results also found that the ranking of economic and ecological subcategory words changed among different IRs, with ecological activities emerging in the IR-III. However, surprisingly, the great technological innovations of the four IRs did not affect human preferences. In contrast, historical events affect the fluctuations of social transitions, such as World War I and II, economic crises, oil crises, international environmental conferences, climate conventions, biodiversity conventions, the "Great Leap Forward" in 1958, and the “Cultural Revolution" from 1966 to 1976 in China (\cite{Liu1}). However, these types of events represent concentrated outbreaks of contradictions in the process of social-economic and ecological development. The fluctuations were not impacted by external events such as volcanoes or earthquakes (\cite{Schot}). In other words, those fluctuations in the trajectory of economic and ecological words were controlled by endogenous forces of society.

\subsection*{4.2 Trajectories tend to constant oscillations}

Economic and ecological systems are interlinked complex adaptive systems (CAS, \cite{Levin}). On their trajectory, the fluctuation cycle and amplitude followed their own rhythms. The fluctuation periods of the economic words continuously have shortened to 25-30 years in recent years, and the ecological cycle shortened to about 10 years in the two countries. This phenomenon is also endogenous, as it is regardless of the impact of various events during this period. The shortening of the ecological cycle even conforms to a power law, which emphasizes the self-organizing behaviour of human society. In addition, many cultural processes, such as the lifespan of albums, the turnover of music charts, and human forgetting, have also been shortening (\cite{Michel}; \cite{Schneider}). The cycles of economic and ecological activities are similar to the "short cycles" related to human behaviour (\cite{Schot}). For example, the cycle of finance is 16 years (\cite{Borio}), the fluctuation period of the price-to-earnings ratio of a stock is about 30 years (\cite{Shiller}) and the popularity cycles of musical instruments and fashion are about 30 years (\cite{Mauch}; \cite{Kim}). Those are all close to the activity cycle of a generation in human society, which is 20-25 years (\cite{Axelrad}). These findings suggest that the “psychological cycle” (\cite{Gangestad}) also applies to economic and ecological activities. Periodicity also allows us to predict and regulate social behaviour in the future.

The amplitude of fluctuations of both economic and ecological words has also decreased continuously over time, similar to the damped vibration in physics. This means that society generated fluctuations in activities through negative feedback. However, the amplitude approached stability in the later stage, and the amplitudes of the last 3-4 cycles of the two kinds of words in both countries remained stable at CV around 20\% since 1960. This stable amplitude is similar to the simple harmonic motion in physics (Gilbert, 2022). Most trajectories of a dynamical system do not converge to the rest points (\cite{Levin}; \cite{Roy}). Economic and ecological activities, as a dynamic CAS, generated multiple equilibria that could be stable or unstable. The findings imply the similarity between social and physical systems. Maintaining the constant oscillation motion in social consciousness requires a force, which should be human preference. Cultural history and behavioural ecology assume that when rigid demands are met, the preference for elastic factors increases (\cite{Thiermann}). For example, with economic improvement, the love in narrative novels (\cite{Baumard}) and the biodiversity of urban green spaces (\cite{Leong}) increase, while the tolerance for environmental pollution decreases (\cite{Liu}). The fluctuation variation of ecological words is larger than that of economic words in the early period, indicating that economic activity is more rigid than ecological activity. The recent CV of ecological activities was close to and even lower than that of economic activities. This means that society's demand for ecological development is becoming a rigid element. In addition, in statistics, the CV of 20\% belongs to moderate variation, indicating that the activities in the two countries have developed to be both energetic and stable.

The turnover rate of economic and ecological words increases with a constant speed as the interval years, and it is similar to a molecular clock. For example, the molecular clock was \(1.2 \times 10^{-9}\) subs \(site^{-1}\)  \(yr^{-1}\) between humans and chimpanzees (Li and Tanimura, 1987). Since 1960, the turnover rates have stabilized at 60-80 species per year in the UK and China, which we call a “cultural clock”. This indicates that the rhythm of economic and ecological activities and the related social consciousness have trended towards stabilization. In other words, they have entered a mature stage of development. The ecological activities in China are not yet mature.

The Simpson indices increased to near 1, which is the high limit despite the great difference between the economic and ecological word species in the early stage in both countries. The highest diversity means the maximum differentiation of ecological communities (\cite{Morris}), which helps to promote stability (\cite{Isbell}). Social activities have reached a relatively stable state in both countries with different industrialization levels, which confirms cultural dissemination in social transition.
The turnover distribution, it can be predicted that the dominant category of economic words, such as financial related-words, and ecological words, such as environmental protection related-words, will maintain an advantage in the UK. The high-frequency economic words in China have rapidly changed, and financial-related words, currently occupying the top spot, may soon be surpassed by commercial-related words. The high-frequency ecological words in the future may still be environmental protection-related words. This indicates that the vitality of ecological activities is still weak and needs to be improved in future sustainability.

\subsection*{4.3 Sustainability is achievable but unstable}

There is asynchronism and uncertainty in the co-evolution of ecological and economic activities (\cite{Levin}). The relationship between economic and ecological words exhibited the Z-mode in both countries. It tended to synergistically increase with 2/3 of the industrialization, performed as increasing of both economic and ecological words, while trade-off with only one side increasing or declining, or both declines in the rest of the time. The multiple cycles of continuous synergy and trade-offs during industrialization are challenges to sustainable social transition (\cite{Bettencourt}). Although ecological awareness has increased with economic development, the occurrence of ecological words is still only 1/20 of that of economic words in the UK, and the continuous improvement of ecology is still a challenge. In China, ecological activities are in the process of catching up, so the proportion has reached about 1/4. In addition, ecological development cannot be separated from economic foundations (\cite{Frank}). If there is a partial or short-term economic recession, the economic-ecological synergy will be destroyed. For example, during the Great Leap Forward movement (1958-1960) in China, a large number of trees were chopped down to make steel; the 2022 European energy crisis led to setbacks in the "carbon neutrality" initiative, and some people even logging trees in parks as winter fuel. This Z-shaped pattern indicates the complex co-evolution behaviour. It also indicates a serious problem, that is, even if societies have developed, sustainability is still vulnerable, with insufficient robustness and a significant risk of reversal.

We take the occurrence of negative ecological words as an indicator of environmental degradation of the EKC hypothesis, which assumes that ecological degradation undergoes an inverted U-shaped mode with the economic level of society (\cite{Zhang2}; \cite{Anser}). We find that after the 1960s, when the economic words in the UK entered the stabilization stage, the negative ecological words performed 4 inverted U-shaped curves in response to GDP per capita within 50,000 USD. The negative ecological words in China performed 2 inverted U-shaped curves in response to GDP per capita within 10,000 USD. We named the multiple inverted U-shaped curves the M-shaped curves, which is a new type beyond the inverted U-shape (\cite{Zhang2}), the N-shaped (\cite{Danish}), and the inverted V-shaped (\cite{Yu}) curves for the EKC model.

This M model means that even after the transition, when a society enters the sustainable ‘Red Loop’ defined by \cite{Cumming2}, it can still be disrupted. A red loop means that social demands have been decoupled from natural ecosystems and that society continuously regulates supply and distribution problems through feedback via technological innovations to achieve sustainable development (\cite{Chang}; \cite{Bettencourt}). If the feedback fails, a society will enter a ‘Red Trap’ again. Therefore, social sustainability requires continuous reconstruction in the co-evolution of ecological and economic activities. At present, the inverted U-shaped curve occurs with GDP per capita within 10,000 USD in China, which corresponds to the GDP per capita range of the first inverted U-shaped curve in the UK. It can be inferred that the synergy between economic and ecological development in China may also generate more fluctuations with future economic development. Fortunately, despite China having many serious ecological problems in its early stages of development (\cite{Liu1}), the ecological awareness is continuously increasing. It exhibits a similar co-evolution model with that in the UK. In sum, the findings can help to speculate that, under the influence of cultural dissemination and endogenous forces, latecomers will not only follow the general pattern similar to the original countries but also exhibit stronger sustainable transition vitality to catch up with industrialized countries.

\subsection*{4.4 Limitations}
This is a cross-cultural study, and there is currently no effective tool to address differences in language, politics, or customs among different societies. However, a recent study has shown that the information rates conveyed by different languages are similar (about 39 bits/s, \cite{Coupé}, which suggests the comparability of cross-language information. The quantification of social culture is a daunting task (\cite{Gao}), as human economic and ecological activities involve a wide range and complexity, making it difficult to fully cover the construction of a targeted library. We need to supplement the targeted words more comprehensively by building an active ontology and consulting the experts as much as possible (see methods). Despite these limitations, the use of newspapers to quantify the evolutionary trajectory of social consciousness and human activities is still an important intellectual discourse, and has higher timeliness than academic journals and books (\cite{Lansdall-Welfare}). In this study, we have attempted to select a symmetrical library for the two countries. Considering the diachronic and dynamic differences in corpora, we also used other independent corpora for cross-supplementation (see methods). Future work may link word frequency based on the corpus with actual biophysical data (ecosystem services, human development index) and demonstrate how to expand our research to more areas of social transformation, such as art, industry, religion, and technology.

\section*{5 Conclusions}
Using culturomics, we find that cultural dissemination can promote the transition of a latecomer society, but the receiver still needs self-development. Economic and ecological activities entered a mega trend of synergistic growth after IR-III. However, the co-evolution of them experienced multiple short-term reverses and exhibited a Z-mode. Although the public consciousness of sustainability has improved and the rigidity of ecological elements in human well-being has increased, the economic elements, which is closer to personal interests, is still dominant. The third M-shaped mode indicates that social sustainability is not robust enough and requires periodic reconstruction in the co-evolution.

We discovered interesting near-physical dynamic features. These findings also contribute to a deeper understanding of the physical laws and evolutionary mechanisms of collective behaviour in human society.

The turnover rates acted as a ‘cultural clock’ for the stabilized speed across years, while the pace of ecological activity changes was slower than that of economic changes. This means that in social management, it is necessary to identify these pulses related to sustainability and prevent and optimize them.

%\clearpage

\section*{Acknowledgments}

This work was financially supported by the National Natural Science Foundation of China (Grant No.31770434). We thank Professor Ming Chen and Danqing Zhang for the comments and contributions.

\nolinenumbers

%This is where your bibliography is generated. Make sure that your .bib file is actually called library.bib

%This defines the bibliographies style. Search online for a list of available styles.
\bibliographystyle{plainnat} % or try abbrvnat or unsrtnat
\bibliography{library} % refers to example.bib

\begin{thebibliography}{57}
\providecommand{\natexlab}[1]{#1}
\providecommand{\url}[1]{\texttt{#1}}
\expandafter\ifx\csname urlstyle\endcsname\relax
  \providecommand{\doi}[1]{doi: #1}\else
  \providecommand{\doi}{doi: \begingroup \urlstyle{rm}\Url}\fi

\bibitem[Acerbi and Bentley(2014)]{Acerbi}
A.~Acerbi and R.~A. Bentley.
\newblock Biases in cultural transmission shape the turnover of popular traits.
\newblock \emph{Evol. Hum. Behav}, 35:\penalty0 228--236, 2014.
\newblock URL \url{http://biorxiv.org/content/early/2016/04/07/044990}.

\bibitem[Ali(2023)]{Ali}
J.~Ali.
\newblock Convergence innovation for sustainable development: Unraveling post-covid dynamics for a resilient future.
\newblock \emph{Sustain. Dev.}, 35:\penalty0 1--11, 2023.
\newblock URL \url{https://doi.org/10.1002/sd.2810}.

\bibitem[Anderson et~al.(2021)Anderson, Elsen, Hughes, Tonietto, Bletz, Gill, Holgerson, Kuebbing, MacKenzie, Meek, and Veríssimo]{Anderson}
S.C. Anderson, P.R. Elsen, B.B. Hughes, R.K. Tonietto, M.C. Bletz, D.A. Gill, M.A. Holgerson, S.E Kuebbing, M.C. MacKenzie, M.H. Meek, and D.~Veríssimo.
\newblock Trends in ecology and conservation over eight decades.
\newblock \emph{Frontiers in ecology and the environment}, 19\penalty0 (5):\penalty0 274--282, 2021.
\newblock URL \url{https://doi.org/10.1002/fee.2320}.

\bibitem[Anser et~al.(2021)Anser, Usman, Godil, Shabbir, Sharif, Tabash, and Lopez]{Anser}
M.~K. Anser, Muhammad Usman, Danish~I. Godil, Malik~S. Shabbir, Arshian Sharif, Mosab~I. Tabash, and Lydia~B. Lopez.
\newblock Does globalization affect the green economy and environment? the relationship between energy consumption, carbon dioxide emissions, and economic growth.
\newblock \emph{Environmental science and pollution research international}, 28\penalty0 (37):\penalty0 51105--51118, 2021.
\newblock URL \url{https://doi.org/10.1007/s11356-021-14243-4}.

\bibitem[Atenstaedt(2021)]{Atenstaedt}
R.~L. Atenstaedt.
\newblock Word cloud analysis of historical changes in the subject matter of public health practice in the united kingdom.
\newblock \emph{Public health (London)}, 197:\penalty0 39--41, 2021.
\newblock URL \url{https://doi.org/10.1007/s11356-021-14243-4}.

\bibitem[Axelrad and Luski(2022)]{Axelrad}
H.~Axelrad and I.~Luski.
\newblock Actual retirement age: A european cross-country analysis.
\newblock \emph{Ageing international}, 47\penalty0 (3):\penalty0 534--558, 2022.
\newblock URL \url{https://doi.org/10.1007/s11356-021-14243-4}.

\bibitem[Baselga and Orme(2012)]{Baselga}
Andrés Baselga and C.~D.~L. Orme.
\newblock betapart: an r package for the study of beta diversity.
\newblock \emph{Methods in ecology and evolution}, 3\penalty0 (5):\penalty0 808--812, 2012.
\newblock URL \url{https://doi.org/10.1111/j.2041-210X.2012.00224.x}.

\bibitem[Baumard et~al.(2022)Baumard, Huillery, Hyafil, and Safra]{Baumard}
Nicolas Baumard, Elise Huillery, Alexandre Hyafil, and Lou Safra.
\newblock The cultural evolution of love in literary history.
\newblock \emph{Nature human behaviour}, 6\penalty0 (4):\penalty0 506--522, 2022.
\newblock URL \url{https://doi.org/10.1038/s41562-022-01292-z}.

\bibitem[Bentley et~al.(2012)Bentley, Garnett, O'Brien, and Brock]{Bentley}
R.~A. Bentley, Philip Garnett, Michael~J. O'Brien, and William~A. Brock.
\newblock Word diffusion and climate science.
\newblock \emph{PloS one}, 7\penalty0 (11):\penalty0 e47966, 2012.
\newblock URL \url{https://doi.org/10.1371/journal.pone.0047966}.

\bibitem[Bettencourt et~al.(2007)Bettencourt, Lobo, Helbing, Kühnert, and West]{Bettencourt}
Luís M.~A. Bettencourt, José Lobo, Dirk Helbing, Christian Kühnert, and Geoffrey~B. West.
\newblock Growth, innovation, scaling, and the pace of life in cities.
\newblock \emph{Proceedings of the National Academy of Sciences}, 104\penalty0 (17):\penalty0 7301--7306, 2007.
\newblock URL \url{www.pnas.org/cgi/content/full/0610172104/DC1}.

\bibitem[Boivin and Crowther(2021)]{Boivin}
Nicole Boivin and Alison Crowther.
\newblock Mobilizing the past to shape a better anthropocene.
\newblock \emph{Nature ecology and evolution}, 5\penalty0 (3):\penalty0 273--284, 2021.
\newblock URL \url{https://doi.org/10.1038/s41559-020-01361-4}.

\bibitem[Borio(2012)]{Borio}
Claudio Borio.
\newblock The financial cycle and macroeconomics: What have we learnt?
\newblock \emph{J. Bank Financ.}, 45:\penalty0 182--198, 2012.
\newblock URL \url{http://dx.doi.org/10.1016/j.jbankﬁn.2013.07.031}.

\bibitem[Chang et~al.(2021)Chang, Ge, Wu, Du, Pan, Yang, Ren, Helio, Mao, Cheong, Qu, Fan, Min, Peng, and Meyerson]{Chang}
J.~Chang, Y.~Ge, Z.~Wu, Y.~Du, K.~Pan, G.~Yang, Y.~Ren, M.P. Helio, F.~Mao, K.~H. Cheong, Z.~Qu, X.~Fan, Y.~Min, C.~H. Peng, and L.A. Meyerson.
\newblock Modern cities modelled as “super-cells” rather than multicellular organisms: Implications for industry, goods and services.
\newblock \emph{Bioessays}, 43\penalty0 (2100041), 2021.
\newblock URL \url{https://onlinelibrary.wiley.com/doi/10.1002/bies.202100041}.

\bibitem[Coupé et~al.(2019)Coupé, Oh, Dediu, and Pellegrino]{Coupé}
Christophe Coupé, Yoon Oh, Dan Dediu, and François Pellegrino.
\newblock Different languages, similar encoding efficiency: Comparable information rates across the human communicative niche.
\newblock \emph{Science advances}, 5\penalty0 (9):\penalty0 2594--2594, 2019.
\newblock URL \url{https://doi.org/10.1126/sciadv.aaw2594}.

\bibitem[Cumming and von Cramon-Taubadel(2018)]{Cumming1}
Graeme~S. Cumming and Stephan von Cramon-Taubadel.
\newblock Linking economic growth pathways and environmental sustainability by understanding development as alternate social–ecological regimes.
\newblock \emph{Proceedings of the National Academy of Sciences - PNAS}, 115\penalty0 (38):\penalty0 9533--9538, 2018.
\newblock URL \url{https://doi.org/10.1073/pnas.1807026115}.

\bibitem[Cumming et~al.(2014)Cumming, Buerkert, Hoffmann, Schlecht, Von Cramon-Taubadel, and Tscharntke]{Cumming2}
Graeme~S. Cumming, Andreas Buerkert, Ellen~M. Hoffmann, Eva Schlecht, Stephan Von Cramon-Taubadel, and Teja Tscharntke.
\newblock Implications of agricultural transitions and urbanization for ecosystem services.
\newblock \emph{Nature}, 515\penalty0 (7525):\penalty0 50--57, 2014.
\newblock URL \url{https://doi.org/10.1038/nature13945}.

\bibitem[Danish and Wang(2019)]{Danish}
D.~Danish and Z.~Wang.
\newblock Does biomass energy consumption help to control environmental pollution? evidence from brics countries.
\newblock \emph{The Science of the total environment}, 670:\penalty0 1075--1083, 2019.
\newblock URL \url{https://doi.org/10.1016/j.scitotenv.2019.03.268}.

\bibitem[Dutta et~al.(2015)Dutta, CHATTERJEE, and Madalli]{Dutta}
B.~Dutta, U.~CHATTERJEE, and Devika~P. Madalli.
\newblock Yamo: Yet another methodology for large-scale faceted ontology construction.
\newblock \emph{Journal of knowledge management}, 19\penalty0 (1):\penalty0 6--24, 2015.
\newblock URL \url{https://doi.org/10.1108/JKM-10-2014-0439}.

\bibitem[Frank and Schlenker(2016)]{Frank}
Eyal~G. Frank and W.~Schlenker.
\newblock Balancing economic and ecological goals what are the trade-offs between economic development and ecosystem conservation?
\newblock \emph{Science (American Association for the Advancement of Science)}, 353\penalty0 (6300):\penalty0 651--652, 2016.
\newblock URL \url{https://doi.org/10.1126/science.aaf9697}.

\bibitem[Gangestad et~al.(2019)Gangestad, Dinh, Grebe, Del~Giudice, and Emery~Thompson]{Gangestad}
Steven~W. Gangestad, Tran Dinh, Nicholas~M. Grebe, Marco Del~Giudice, and Melissa Emery~Thompson.
\newblock Psychological cycle shifts redux: Revisiting a preregistered study examining preferences for muscularity.
\newblock \emph{Evolution and human behavior}, 40\penalty0 (6):\penalty0 501--516, 2019.
\newblock URL \url{https://doi.org/10.1016/j.evolhumbehav.2019.05.005}.

\bibitem[Gao et~al.(2019)Gao, Zhang, and Zhou]{Gao}
Jian Gao, Yi~C. Zhang, and Tao Zhou.
\newblock Computational socioeconomics.
\newblock \emph{Physics reports}, 817:\penalty0 1--104, 2019.
\newblock URL \url{https://linkinghub.elsevier.com/retrieve/pii/S0370157319301954}.

\bibitem[Gilbert(2021)]{Gilbert}
P.~U. P.~A. Gilbert.
\newblock \emph{Physics in the Arts, Third Edition}.
\newblock Academic Press, 2021.

\bibitem[Gu et~al.(2023)Gu, Zeng, Chen, Balezentis, and Sapolaite]{Gu}
Jiaxing Gu, Shouzhen Zeng, Wendi Chen, Tomas Balezentis, and Vaida Sapolaite.
\newblock Tracking sustainable development from the dual perspective of environmental and economic performance: A dynamic framework and coupling coordination degree.
\newblock \emph{Sustainable development (Bradford, West Yorkshire, England)}, page 1–14, 2023.
\newblock URL \url{https://doi.org/10.1002/sd.2816}.

\bibitem[Hills et~al.(2019)Hills, Proto, Sgroi, and Seresinhe]{Hills}
Thomas~T. Hills, Eugenio Proto, Daniel Sgroi, and Chanuki~I. Seresinhe.
\newblock Historical analysis of national subjective wellbeing using millions of digitized books (vol 3, pg 1271, 2019).
\newblock \emph{Nature human behaviour}, 3\penalty0 (12):\penalty0 1343--1343, 2019.
\newblock URL \url{https://doi.org/10.1038/s41562-019-0781-5}.

\bibitem[Isbell et~al.(2015)Isbell, Craven, Connolly, Loreau, Schmid, Beierkuhnlein, Bezemer, Bonin, Bruelheide, De~Luca, Ebeling, Griffin, Guo, Hautier, Hector, Jentsch, Kreyling, Lanta, Manning, Meyer, Mori, Naeem, Niklaus, Polley, Reich, Roscher, Seabloom, Smith, Thakur, Tilman, Tracy, Van Der~Putten, Van~Ruijven, Weigelt, Weisser, Wilsey, and Eisenhauer]{Isbell}
Forest Isbell, Dylan Craven, John Connolly, Michel Loreau, Bernhard Schmid, Carl Beierkuhnlein, T.~M. Bezemer, Catherine Bonin, Helge Bruelheide, Enrica De~Luca, Anne Ebeling, John~N. Griffin, Qinfeng Guo, Yann Hautier, Andy Hector, Anke Jentsch, Jürgen Kreyling, Vojtêch Lanta, Pete Manning, Sebastian~T. Meyer, Akira~S. Mori, Shahid Naeem, Pascal~A. Niklaus, H.~W. Polley, Peter~B. Reich, Christiane Roscher, Eric~W. Seabloom, Melinda~D. Smith, Madhav~P. Thakur, David Tilman, Benjamin~F. Tracy, Wim~H. Van Der~Putten, Jasper Van~Ruijven, Alexandra Weigelt, Wolfgang~W. Weisser, Brian Wilsey, and Nico Eisenhauer.
\newblock Biodiversity increases the resistance of ecosystem productivity to climate extremes.
\newblock \emph{Nature (London)}, 526\penalty0 (7574):\penalty0 574--577, 2015.
\newblock URL \url{https://doi.org/10.1038/nature15374}.

\bibitem[K.~H. and B.~H.(2020)]{Kim}
Kim K.~H. and Won B.~H.
\newblock The analysis of fashion trend cycle using big data.
\newblock \emph{J. Korea Convergence Soc.}, 11:\penalty0 113--123, 2020.
\newblock URL \url{https://doi.org/10.15207/JKCS.2020.11.12.113}.

\bibitem[Kates(2011)]{Kates}
Robert~W. Kates.
\newblock What kind of a science is sustainability science?
\newblock \emph{Proceedings of the National Academy of Sciences - PNAS}, 108\penalty0 (49):\penalty0 19449--19450, 2011.
\newblock URL \url{https://doi.org/10.1073/pnas.1116097108}.

\bibitem[Krausmann et~al.(2008)Krausmann, Schandl, and Sieferle]{Krausmann}
Fridolin Krausmann, Heinz Schandl, and Rolf~P. Sieferle.
\newblock Socio-ecological regime transitions in austria and the united kingdom.
\newblock \emph{Ecological economics}, 65\penalty0 (1):\penalty0 187--201, 2008.
\newblock URL \url{https://doi.org/10.1016/j.ecolecon.2007.06.009}.

\bibitem[Lansdall-Welfare et~al.(2017)Lansdall-Welfare, Sudhahar, Thompson, Lewis, Team, Cristianini, and Team]{Lansdall-Welfare}
Thomas Lansdall-Welfare, Saatviga Sudhahar, James Thompson, Justin Lewis, FindMyPast~N. Team, Nello Cristianini, and FindMyPast~Newspaper Team.
\newblock Content analysis of 150 years of british periodicals.
\newblock \emph{Proceedings of the National Academy of Sciences - PNAS}, 114\penalty0 (4):\penalty0 E457--E465, 2017.
\newblock URL \url{https://doi.org/10.1073/pnas.1606380114}.

\bibitem[Leong et~al.(2018)Leong, Dunn, and Trautwein]{Leong}
Misha Leong, Robert~R. Dunn, and Michelle~D. Trautwein.
\newblock Biodiversity and socioeconomics in the city: a review of the luxury effect.
\newblock \emph{Biology letters (2005)}, 14\penalty0 (5):\penalty0 20180082--20180082, 2018.
\newblock URL \url{http://dx.doi.org/10.1098/rsbl.2018.0082}.

\bibitem[Levin and Xepapadeas(2021)]{Levin}
Simon Levin and Anastasios Xepapadeas.
\newblock On the coevolution of economic and ecological systems.
\newblock \emph{Annual review of resource economics}, 13\penalty0 (1):\penalty0 355--377, 2021.
\newblock URL \url{https://doi.org/10.1146/annurev-resource-103020-083100}.

\bibitem[Li and Tanimura(1987)]{Li}
W.~Li and M.~Tanimura.
\newblock The molecular clock runs more slowly in man than in apes and monkeys.
\newblock \emph{Nature (London)}, 326\penalty0 (6108):\penalty0 93--96, 1987.
\newblock URL \url{https://doi.org/10.1038/326093a0}.

\bibitem[Liu and Diamond(2005)]{Liu1}
J.~Liu and J.~Diamond.
\newblock China’s environment in a globalizing world–how china and the rest of the world affect each other.
\newblock \emph{Nature}, 435:\penalty0 1179–1186, 2005.
\newblock URL \url{https://doi.org/10.1038/4351179a}.

\bibitem[Liu and Mu(2016)]{Liu}
X.~Liu and R.~Mu.
\newblock Public environmental concern in china: Determinants and variations.
\newblock \emph{Global environmental change}, 37:\penalty0 116--127, 2016.
\newblock URL \url{http://dx.doi.org/10.1016/j.gloenvcha.2016.01.008}.

\bibitem[Madin et~al.(2008)Madin, Bowers, Schildhauer, and Jones]{Madin}
J.~S. Madin, S.~Bowers, M.~P. Schildhauer, and M.~B. Jones.
\newblock Advancing ecological research with ontologies.
\newblock \emph{Trends in ecology and evolution (Amsterdam)}, 23\penalty0 (3):\penalty0 159--168, 2008.
\newblock URL \url{https://doi.org/10.1016/j.tree.2007.11.007}.

\bibitem[Mao et~al.(2021)Mao, Hu, Liu, and Crittenden]{Mao}
G.~Mao, H.~Hu, X.~Liu, and N.~Crittenden, J.and~Huang.
\newblock A bibliometric analysis of industrial wastewater treatments from 1998 to 2019.
\newblock \emph{Environ. Pollut. (1987)}, 275\penalty0 (115785), 2021.
\newblock URL \url{https://doi.org/10.1016/j.envpol.2020.115785}.

\bibitem[Markard(2020)]{Markard}
Jochen Markard.
\newblock The life cycle of technological innovation systems.
\newblock \emph{Technological forecasting and social change}, 153\penalty0 (119407), 2020.
\newblock URL \url{https://doi.org/10.1016/j.techfore.2018.07.045}.

\bibitem[Mauch et~al.(2015)Mauch, MacCallum, Levy, and Leroi]{Mauch}
Matthias Mauch, Robert~M. MacCallum, Mark Levy, and Armand~M. Leroi.
\newblock The evolution of popular music: Usa 1960–2010.
\newblock \emph{Royal Society open science}, 2\penalty0 (5):\penalty0 150081--150081, 2015.
\newblock URL \url{https://doi.org/10.1098/rsos.150081}.

\bibitem[Meng et~al.(2021)Meng, Guo, Guo, Lee, Liu, and Wang]{Meng}
F.~Meng, J.~Guo, Z.~Guo, J.~C.~K. Lee, G.~Liu, and N.~Wang.
\newblock Urban ecological transition: The practice of ecological civilization construction in china.
\newblock \emph{The Science of the total environment}, 755, 2021.
\newblock URL \url{https://doi.org/10.1016/j.scitotenv.2020.142633}.

\bibitem[Michel et~al.(2011)Michel, Shen, Aiden, Veres, Gray, Pickett, Hoiberg, Clancy, Norvig, Orwant, Pinker, Nowak, Aiden, Team, and Team]{Michel}
Jean-Baptiste Michel, Yuan~K. Shen, Aviva~P. Aiden, Adrian Veres, Matthew~K. Gray, Joseph~P. Pickett, Dale Hoiberg, Dan Clancy, Peter Norvig, Jon Orwant, Steven Pinker, Martin~A. Nowak, Erez~L. Aiden, The Google~Books Team, and Google~Books Team.
\newblock Quantitative analysis of culture using millions of digitized books.
\newblock \emph{Science (American Association for the Advancement of Science)}, 331\penalty0 (6014):\penalty0 176--182, 2011.
\newblock URL \url{https://doi.org/10.1126/science.1199644}.

\bibitem[Miller et~al.(2012)Miller, Gomberg-Maitland, and Humbert]{Miller}
Dave~P. Miller, Mardi Gomberg-Maitland, and Marc Humbert.
\newblock Survivor bias and risk assessment.
\newblock \emph{The European respiratory journal}, 40\penalty0 (3):\penalty0 530--532, 2012.
\newblock URL \url{https://doi.org/10.1183/09031936.00094112}.

\bibitem[Morris et~al.(2014)Morris, Caruso, Buscot, Fischer, Hancock, Maier, Meiners, Müller, Obermaier, Prati, Socher, Sonnemann, Wäschke, Wubet, Wurst, and Rillig]{Morris}
E.~K. Morris, Tancredi Caruso, François Buscot, Markus Fischer, Christine Hancock, Tanja~S. Maier, Torsten Meiners, Caroline Müller, Elisabeth Obermaier, Daniel Prati, Stephanie~A. Socher, Ilja Sonnemann, Nicole Wäschke, Tesfaye Wubet, Susanne Wurst, and Matthias~C. Rillig.
\newblock Choosing and using diversity indices: insights for ecological applications from the german biodiversity exploratories.
\newblock \emph{Ecology and evolution}, 4\penalty0 (18):\penalty0 3514--3524, 2014.
\newblock URL \url{https://doi.org/10.1002/ece3.1155}.

\bibitem[O'Brien(2018)]{O'Brien}
Geoff O'Brien.
\newblock Cities - good for the environment?
\newblock \emph{International journal of environmental studies}, 75\penalty0 (1):\penalty0 16--28, 2018.
\newblock URL \url{https://doi.org/10.1080/00207233.2017.1392767}.

\bibitem[Ouyang et~al.(2023)Ouyang, Chen, Yang, Ren, Yu, Chang, and Ge]{Ouyang}
Yan Ouyang, Yi~Chen, Guofu Yang, Yuan Ren, Mingjian Yu, Jie Chang, and Ying Ge.
\newblock Homogenization of trees in urban green spaces along the moisture gradient in china.
\newblock \emph{Urban forestry and urban greening}, 83, 2023.
\newblock URL \url{https://doi.org/10.1016/j.ufug.2023.127892}.

\bibitem[Palmer et~al.(2004)Palmer, Bernhardt, Chornesky, Collins, Dobson, Duke, Gold, Jacobson, Kingsland, Kranz, Mappin, Martinez, Micheli, Morse, Pace, Pascual, Palumbi, Reichman, Simons, Townsend, and Turner]{Palmer}
Margaret Palmer, Emily Bernhardt, Elizabeth Chornesky, Scott Collins, Andrew Dobson, Clifford Duke, Barry Gold, Robert Jacobson, Sharon Kingsland, Rhonda Kranz, Michael Mappin, M.~L. Martinez, Fiorenza Micheli, Jennifer Morse, Michael Pace, Mercedes Pascual, Stephen Palumbi, O.~J. Reichman, Ashley Simons, Alan Townsend, and Monica Turner.
\newblock Ecology for a crowded planet.
\newblock \emph{Science (American Association for the Advancement of Science)}, 304\penalty0 (5675):\penalty0 1251--1252, 2004.
\newblock URL \url{https://doi.org/10.1126/science.1095780}.

\bibitem[Roy et~al.(2023)Roy, Chowdhury, Kundu, Sar, Banerjee, Rakshit, Mali, Perc, and Ghosh]{Roy}
S.~Roy, S.~N. Chowdhury, S.~Kundu, G.~K. Sar, J.~Banerjee, B.~Rakshit, R.~C. Mali, M.~Perc, and D.~Ghosh.
\newblock Time delays shape the eco-evolutionary dynamics of cooperation. sci. rep. 13, 14331. https://doi.org/10.1038/s41598-023-41519-1.
\newblock \emph{Royal Society open science}, 6\penalty0 (8):\penalty0 190944--190944, 2023.
\newblock URL \url{https://doi.org/10.1098/rsos.190944}.

\bibitem[Schneider and Gros(2019)]{Schneider}
Lukas Schneider and Claudius Gros.
\newblock Five decades of us, uk, german and dutch music charts show that cultural processes are accelerating.
\newblock \emph{Royal Society open science}, 6\penalty0 (8):\penalty0 190944--190944, 2019.
\newblock URL \url{https://doi.org/10.1098/rsos.190944}.

\bibitem[Schot and Kanger(2018)]{Schot}
Johan Schot and Laur Kanger.
\newblock Deep transitions: Emergence, acceleration, stabilization and directionality.
\newblock \emph{Research policy}, 47\penalty0 (6):\penalty0 1045--1059, 2018.
\newblock URL \url{https://doi.org/10.1016/j.respol.2018.03.009}.

\bibitem[Shiller(2016)]{Shiller}
Robert~J. Shiller.
\newblock \emph{Irrational exuberance}.
\newblock Princeton University Press, Princeton, revis and expand third edition, 2016.

\bibitem[Skrebyte et~al.(2016)Skrebyte, Garnett, and Kendal]{Skrebyte}
Agne Skrebyte, Philip Garnett, and Jeremy~R. Kendal.
\newblock Temporal relationships between individualism–collectivism and the economy in soviet russia: A word frequency analysis using the google ngram corpus.
\newblock \emph{Journal of cross-cultural psychology}, 47\penalty0 (9):\penalty0 1217--1235, 2016.
\newblock URL \url{https://doi.org/10.1177/0022022116659540}.

\bibitem[Thiermann and Sheate(2020)]{Thiermann}
Ute~B. Thiermann and William~R. Sheate.
\newblock Motivating individuals for social transition: The 2-pathway model and experiential strategies for pro-environmental behaviour.
\newblock \emph{Ecological economics}, 174, 2020.
\newblock URL \url{https://doi.org/10.1016/j.ecolecon.2020.106668}.

\bibitem[Wei et~al.(2017)Wei, Wei, and Western]{Wei}
Jing Wei, Yongping Wei, and Andrew Western.
\newblock Evolution of the societal value of water resources for economic development versus environmental sustainability in australia from 1843 to 2011.
\newblock \emph{Global environmental change}, 42:\penalty0 82--92, 2017.
\newblock URL \url{http://dx.doi.org/10.1016/j.gloenvcha.2016.12.005}.

\bibitem[Xun et~al.(2016)Xun, Rao, Xiao, and Zang]{Xun}
E.~Xun, G.~Q. Rao, X.~Y. Xiao, and J.~J. Zang.
\newblock Construction of bcc corpus in the context of big data.
\newblock \emph{Corpus Linguist}, 3:\penalty0 93–109, 2016.

\bibitem[Y. and G.(2016)]{Wen}
Wen Y. and Fortier G.
\newblock The visible hand: The role of government in china’s long-awaited industrial revolution.
\newblock \emph{Federal Reserve Bank of St. Louis Review}, 98:\penalty0 189--226, 2016.
\newblock URL \url{https://dx.doi.org/10.20955/wp.2016.016}.

\bibitem[Yu et~al.(2022)Yu, Duan, Li, Peng, Yang, Yan, Bi, and Zou]{Yu}
Yadong Yu, Changqun Duan, Shiyu Li, Changhui Peng, Jian Yang, Kun Yan, Xiaoyi Bi, and Ping Zou.
\newblock Relationship between environmental pollution and economic development in late-developing regions shows an inverted v.
\newblock \emph{The Science of the total environment}, 838, 2022.
\newblock URL \url{https://doi.org/10.1016/j.scitotenv.2022.156295}.

\bibitem[Zhang et~al.(2022)Zhang, Huang, Zhang, Hou, Zhou, Chang, Ge, and Chang]{Zhang1}
Danqing Zhang, Guowen Huang, Jiaen Zhang, Xiaoyu Hou, Tianyi Zhou, Xianyuan Chang, Ying Ge, and Jie Chang.
\newblock The evolution of sustainability ideas in china from 1946 to 2015, quantified by culturomics.
\newblock \emph{Sustainability (Basel, Switzerland)}, 14\penalty0 (10):\penalty0 6038, 2022.
\newblock URL \url{https://doi.org/10.3390/su14106038}.

\bibitem[Zhang et~al.(2015)Zhang, Davidson, Mauzerall, Searchinger, Dumas, and Shen]{Zhang2}
Xin Zhang, Eric~A. Davidson, Denise~L. Mauzerall, Timothy~D. Searchinger, Patrice Dumas, and Ye~Shen.
\newblock Managing nitrogen for sustainable development.
\newblock \emph{Nature (London)}, 528\penalty0 (7580):\penalty0 51--59, 2015.
\newblock URL \url{https://doi.org/10.1038/nature15743}.

\end{thebibliography}
\end{document}